\documentclass{PoS}

\usepackage[percent]{overpic}
\usepackage{color}
\usepackage[utf8]{inputenc} 

\title{ANTARES and IceCube combined search for neutrino point-like and extended sources in the Southern Sky}

\ShortTitle{ANTARES-IceCube combined search for neutrino sources in the Southern Hemisphere}

\author{
The ANTARES and IceCube Collaborations\footnote{For collaboration list, see PoS(ICRC2019) 1177.}\\
\itshape \href{http://antares.in2p3.fr/Collaboration/index2.html}{http://antares.in2p3.fr/Collaboration/index2.html}\\
\itshape \href{http://icecube.wisc.edu/collaboration/authors/icrc19_icecube}{http://icecube.wisc.edu/collaboration/authors/icrc19\_icecube}\\
E-mail: \email{giulia.illuminati@ific.uv.es}
}

\abstract{

The ANTARES neutrino telescope, located in the Mediterranean Sea, and the IceCube neutrino observatory, located at the geographic South Pole, both search for cosmic neutrino events with an instantaneous full-sky field of view. The different characteristics of the two telescopes, in particular the larger instrumented volume of IceCube and the better visibility towards the Southern Sky for neutrino energies below 100 TeV of ANTARES, are exploited in a combined search for point-like and extended sources. The sensitivity to neutrino sources located in the Southern Sky is improved by a factor of $\sim$2 compared to individual studies. The data samples used in this analysis correspond to all track-like and shower-like events from the direction of the Southern Sky which were included in the nine-year ANTARES point-source analysis, combined with the through-going track-like events used in the seven-year IceCube point-source search. In this analysis, the Southern Sky is scanned for possible excesses of neutrinos that might indicate the presence of a source, while the coordinates of predefined candidate neutrinos sources are also evaluated in order to limit the penalty of trials. In addition, special focus is given to the region around the Galactic Centre, treated as an extended neutrinos source, and to the location of the supernova remnant RXJ 1713.7-3946. The result of this combined search for galactic and extra-galactic neutrino sources in the Southern Sky is reported here. No significant evidence of cosmic neutrino sources is found and flux upper limits from the various searches are presented.

\vspace{4mm}
{\bfseries Corresponding author:}
\speaker{Giulia Illuminati}$^{1}$\\
{$^{1}$ \itshape IFIC - Instituto de F\'isica Corpuscular (CSIC - Universitat de Val\`encia) c/ Catedr\'atico Jos\'e Beltr\'an, 2 E-46980 Paterna, Valencia, Spain}\\

}

\FullConference{36th International Cosmic Ray Conference -ICRC2019-\\
		July 24th - August 1st, 2019\\
		Madison, WI, U.S.A.}

\begin{document}

\section{Introduction}\label{sec:intro}

In these proceedings a search for cosmic neutrino sources covering the Southern Sky is presented using data from the ANTARES \cite{ANTARESdetector} and IceCube \cite{IceCubedetector2} neutrino telescopes.
The two telescopes consist of a three-dimensional array of photomultiplier tubes (PMTs) inside a transparent medium -- water or ice --, detecting the Cherenkov radiation induced by the passage of relativistic charged particles. These particles are produced in neutrino interactions with the target medium inside or near the instrumented volume. From the collected Cherenkov light, the energy and direction of the incoming neutrinos are reconstructed.
IceCube detection volume is one cubic kilometer consisting of over 5,000 PMTs spread among 86 vertical strings deployed in the antarctic ice close to the geographic South Pole between depths of 1,450 m and 2,450 m.
ANTARES is located in the Mediterranean Sea, 40 km off the coast of Toulon, France. Twelve lines equipped with a total of 885 PMTs are deployed between 2,010 and 2,470 m below the sea level instrumenting a volume of $\sim$0.01$ \rm{km}^3$.
The advantageous field of view of ANTARES as well as the high statistics provided by IceCube allow for a gain in sensitivity by combining the datasets from both experiments in a joint search for point-like and extended sources in the Southern Sky.
The data samples employed in this search are described in Section~\ref{sec:samples}. The analysis method is presented in Section~\ref{sec:method}, while in Section~\ref{sec:searches} the performed searches and corresponding results are presented. The conclusions are drawn in Section~\ref{sec:concl}.

\section{Data samples}\label{sec:samples}
The data sample used in this analysis corresponds to the events from the Southern Sky included in nine-year and seven-year point-sources searches by ANTARES \cite{lastPSant} and IceCube \cite{lastPSic}, respectively. The ANTARES sample includes both track-like and shower-like events. The IceCube sample consists of only track-like events that traversed the whole detection volume. Tracks are mainly induced by charged current (CC) interactions of muon neutrinos. The relativistic muon produced in the interaction can travel large distances through the medium, with the Cherenkov light being constantly emitted along the track. Neutral current (NC) interactions, and ${\nu_e}$ and ${\nu_\tau}$ CC interactions, produce almost spherical light emission around the interaction vertex. The longer lever arm of the track topology allows for a better reconstruction of the particle direction and therefore for a better median angular resolution, making tracks more suited than showers to search for point-like sources, while showers allow for a better reconstruction of the particle energy. 
The ANTARES data were recorded between early 2007 and the end of 2015. The IceCube data were collected from 2008 to 2015, with the detector operating either in partial configuration with only few strings in operation (samples named IC40, IC59, IC79) or with the full detector installed (samples IC86, 2012-2015). A summary of the data samples is given in Table~\ref{tabSamples}. The selected ANTARES tracks are reconstructed with a median angular resolution better than $0.4^{\circ}$ for energies above 100 TeV, while a median angular resolution of $\sim$3$^{\circ}$ is achieved for showers. 
The median angular resolution of the selected IceCube through-going track events at a neutrino energy of 1 TeV is about $1^{\circ}$ and improves to below $0.4^{\circ}$ for energies above 1 PeV.

\setlength{\tabcolsep}{.3em}
\begin{table*}[h!]
    \caption{\footnotesize Characteristic features of the ANTARES and IceCube samples employed in this search. Only events from the Southern Hemisphere (last column) have been selected for the present analysis.
        \label{tabSamples}
    \medskip}
    \resizebox{\linewidth}{!}{
        \begin{minipage}{1.\textwidth} 
        \centering
 {\def\arraystretch{1.}
    \footnotesize
    \begin{tabular}{ccc}
\hline
ANTARES sample & Livetime [days] & \# of events  \\\hline 
Tracks & 2415 & 5807 \\
Showers & 2415 & 102 \\\hline \hline
IceCube sample & Livetime [days] & \# of events  \\\hline 
IC40 & 376 & 22779 \\
IC59 & 348 & 64257 \\
IC79 & 316 & 44771 \\
IC86 & 333 & 74931 \\
2012-2015 & 1058 & 119231\\\hline

\end{tabular}
      }
      \end{minipage}
    }
\end{table*}

\section{Analysis method}\label{sec:method}

A maximum likelihood estimation is performed to identify clusters of cosmic neutrinos from point-like and extended sources over the background of randomly distributed atmospheric neutrinos and muons.
The used likelihood is defined as

\small
\begin{equation} \label{eq:lik}
   L(n_s, \gamma) = \prod_{j = 1}^{7} \prod_{i = 1}^{ N^{j}} \Big[ \frac{n_s^{j}}{N^{j}} {S}^{ j}_{i}(\gamma) + \Big( 1 - \frac{n_s^{j}}{N^{j}} \Big) {B}^{j}_{i} \Big],
\end{equation} 
\normalsize

\noindent where $n_s$ and $\gamma$ are the unknown total number of signal events and signal spectral index, assuming a single power-law spectrum for the source. $S^j_i$ and $B^j_{i}$ are the values of the signal and background probability density functions (PDFs) for event $i$ in sample $j$. $N^{j}$ is the total number of data events in the $j$ sample, while $n_s^{j}$ is the unknown number of signal events in sample $j$. The signal and background PDFs are given by the product of a directional and an energy term. The definition of these PDFs is slightly different for the ANTARES and the IceCube samples and can be found in \cite{lastPSant} and \cite{lastPSic}, respectively. For the IceCube samples, the spatial PDF is given by a 2-dimensional Gaussian, $P_{\rm{space}}^{\rm{IC}} = \rm{exp}(-\Delta\Psi^2/2\sigma^2)/(2\pi\sigma^2)$, with $\Delta\Psi$ being the angular distance of the event from the source and $\sigma$ being the angular error estimate of the event. When searching for spatially extended sources, the value of $\sigma$ is replaced with $\sigma_{eff} = \sqrt{\sigma^2 + \sigma_{s}^2}$, where $\sigma_{s}$ is the extension of the source assuming a Gaussian profile. For the ANTARES samples, a parameterisation of the point spread function, which describes the distribution of signal events around the location of a point-like source, is used as spatial signal PDF. It is defined as the PDF to reconstruct an event at a given angular distance from its original direction, which is derived from Monte Carlo simulations. The original direction of the events is given by the location of the source in the case of a point-source hypothesis. For extended sources, the PSF is built assuming that the original direction of the events is distributed according to a Gaussian profile around the source location with standard deviation given by the assumed source extension $\sigma_s$. In the likelihood maximisation, the values of the number of signal events $n_s$ ($n_s \geq 0.001$) and the signal spectral index $\gamma$ ($1 \leq \gamma \leq 4$) are being optimised. Moreover, the position of the source is either kept fixed or fitted within specific limits depending on the type of search (see Section \ref{sec:searches}). 
In order to determine the significance of a cluster (pre-trial p-value), a test statistic is computed  as

\small
\begin{equation}
     Q = \log  L(\hat{n}_s, \hat{\gamma}) - \log  L(n_s = 0),
    \label{eq:teststat}
\end{equation}
\normalsize
where $\hat{n}_s$ and $\hat{\gamma}$ are the best-fit values from the likelihood maximisation. When many directions in the sky are investigated, a trial correction has to be applied when estimating the significance of the observation. To this purpose, the observed $Q$ is compared to the test statistic distribution obtained when performing the same analysis on many background-only pseudo-experiments -- pseudo-data sets of data randomised in time to eliminate any clustering. The fraction of $Q$ values which are larger than the observed $Q$ gives the trial-corrected significance (post-trial p-value) of the observation.

\section{Searches and results}\label{sec:searches}

Five different searches for cosmic neutrino sources are performed.
The details of each approach together with the respective results are presented below. 

\subsection{Southern-sky search}\label{sec:sss}

\begin{figure*}[ht]
\centering
\begin{overpic}[width=0.7\textwidth]{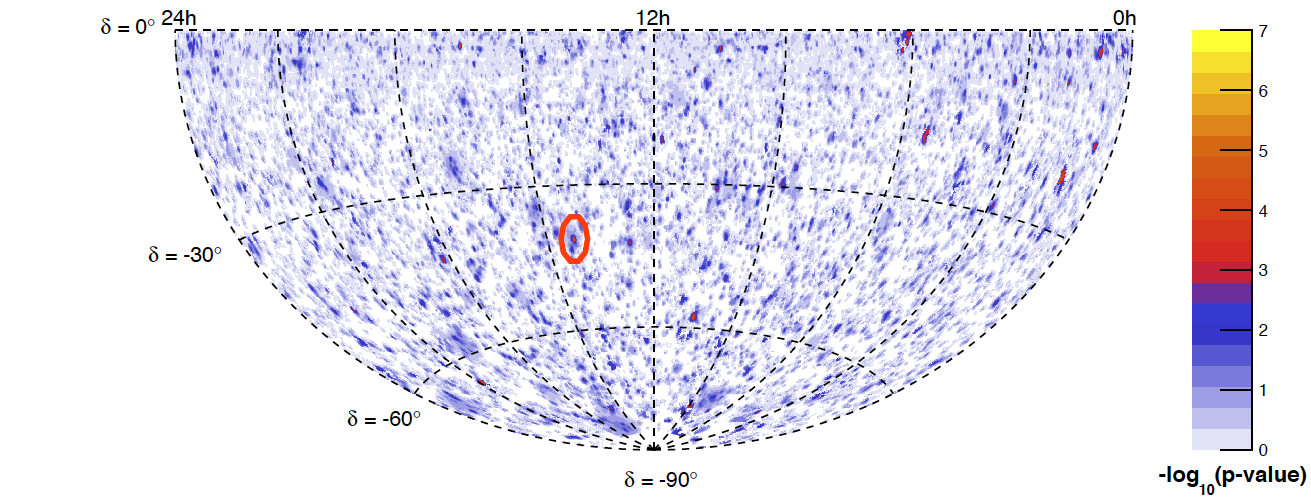}
     \put (65,2.7){\color{black} PRELIMINARY}
    \end{overpic}

\caption{Sky map in equatorial coordinates of the pre-trial p-values obtained in the Southern-sky search for the point-like source hypothesis. The red contour indicates the location of the most significant cluster.}
\label{fig:MapFSS}
\end{figure*}

In this search, a scan of the Southern Sky is performed to find an excess of signal events with respect to the expected background. To this purpose, the $Q$-value defined in equation~(\ref{eq:teststat}) is evaluated in steps of $1^\circ \times 1^\circ$ over the whole scanned region, with the location of the fitted cluster being left free to vary within these boundaries. The results of the searches for the different investigated source extensions $\sigma_s$ ($\sigma_s= 0.0^\circ$ corresponds to the point-like hypothesis) are presented in Table~\ref{tab:FSStable}. The largest excess with respect to the background hypothesis is found at equatorial coordinates ($\hat{\alpha}, \hat{\delta}) = (213.2^\circ, -40.8^\circ)$ for a point-like source hypothesis with best-fit $\hat{n}_s= 5.7$ and $\hat{\gamma} = 2.5$. The trial-corrected significance of the cluster is 18\% (0.9$\sigma$ in the one-sided convention). The position of the cluster together with the pre-trial p-values for all the investigated directions for a point-like source hypothesis is shown in Figure \ref{fig:MapFSS}. 

    \setlength{\tabcolsep}{.3em}
\begin{table*}[h!]
    \caption{\footnotesize List of the most significant clusters found when performing the Southern-sky search for different source hypotheses. Reported are the extension $\sigma_s$ of the Gaussian distribution, the best-fit parameters (number of signal events, $\hat{n}_s$, spectral index, $\hat{\gamma}$,  declination, $\hat{\delta}$, right ascension, $\hat{\alpha}$), and p-values.
        \label{tab:FSStable}
    \medskip}
    \resizebox{\linewidth}{!}{
        \begin{minipage}{1.\textwidth} 
        \centering
 {\def\arraystretch{1.}
    \footnotesize
    \begin{tabular}{c|c|c|c|c|c|c}
   \hline
            $\sigma_s$ $[^\circ]$ & $\hat{n}_s$ & $\hat{\gamma}$ & $\hat{\delta} [^\circ]$   &    $\hat{\alpha}[^\circ]$ & pre-trial p-value & post-trial p-value  \\
\hline

0.0 & 5.7 & 2.5 & -40.8 & 213.2 & $1.3 \times 10^{-5}$  & 0.18 \\ 
0.5 & 10.5 & 3.9 & -22.5 & 18.5 & $3.4 \times 10^{-5}$ & 0.31 \\ 
1.0 & 11.6 & 3.8 & -21.9 & 18.4 & $8.9 \times 10^{-5}$ & 0.44 \\ 
2.0 & 20.3 & 3.0 & -40.1 & 274.1 & $2.2 \times 10^{-4}$ & 0.47 \\ 

\end{tabular}
      }
      \end{minipage}
    }
\end{table*}

\subsection{Candidate list search}\label{sec:cls}
\begin{figure*}[ht]
\centering
\begin{overpic}[width=0.43\textwidth]{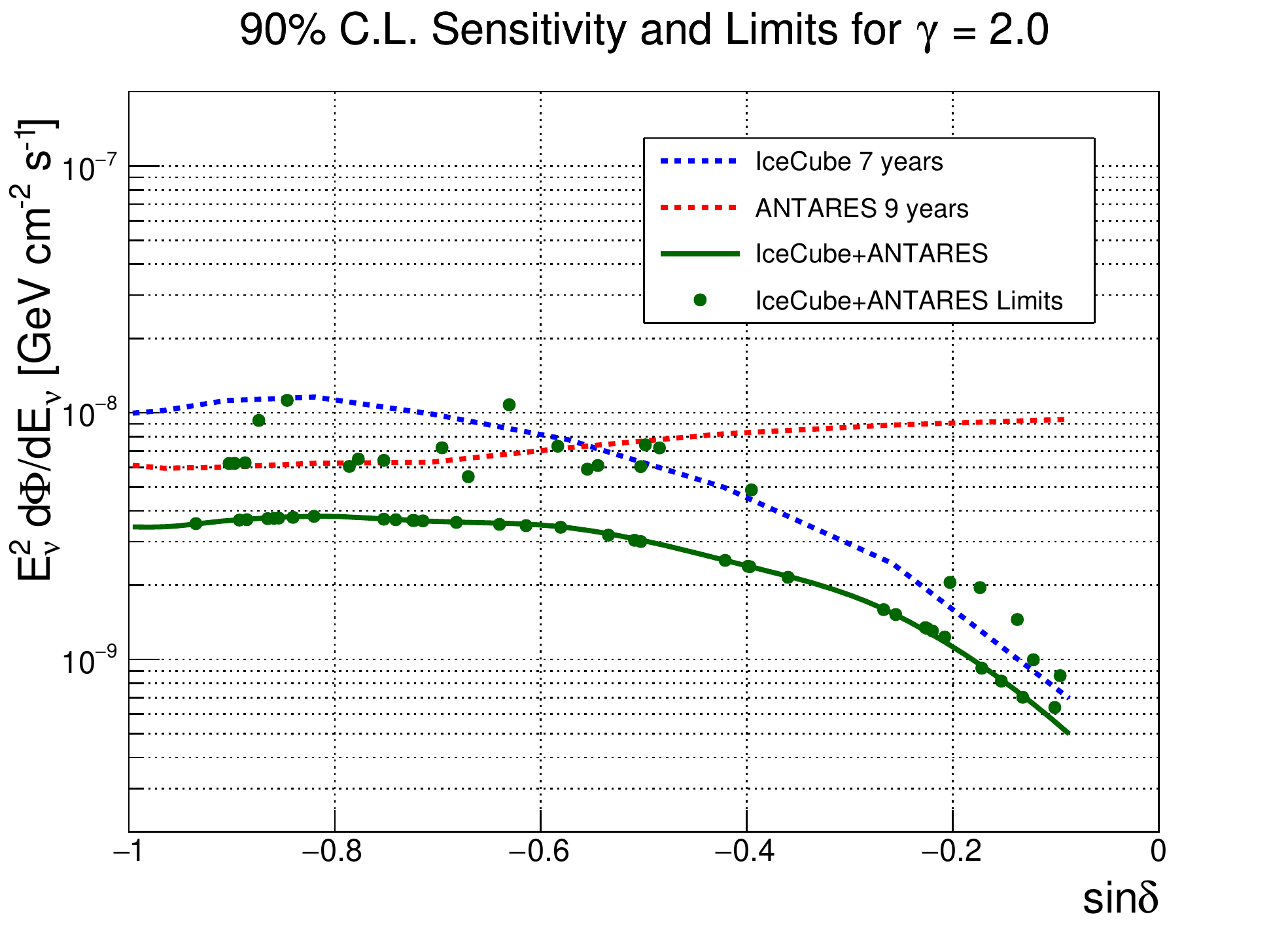}
     \put (11,10.7){\color{black} PRELIMINARY}
    \end{overpic}
    \begin{overpic}[width=0.43\textwidth]{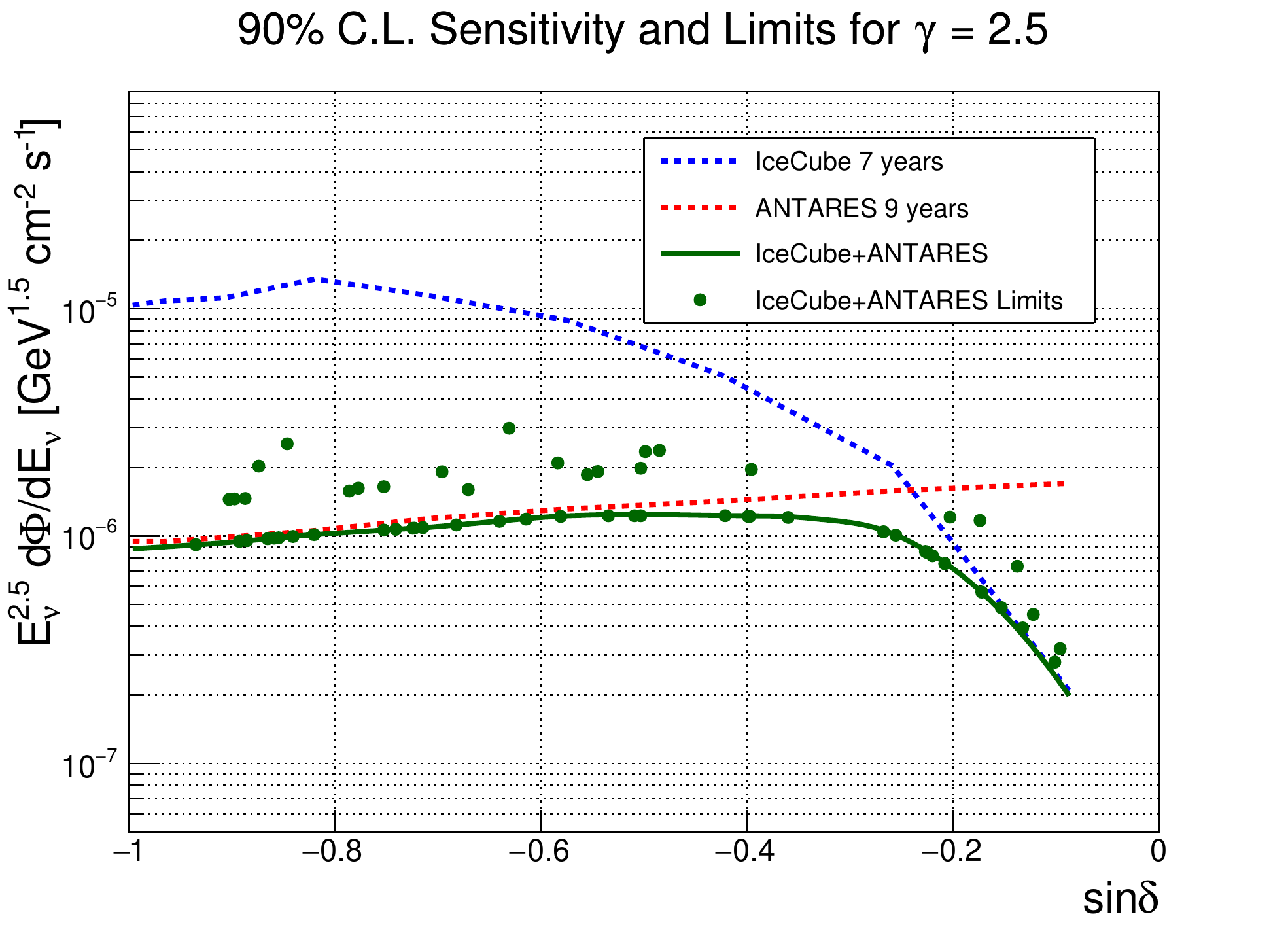}
     \put (11,10.7){\color{black} PRELIMINARY}
    \end{overpic}
\caption{Upper limits at 90\% C.L. on the signal flux from the analysed candidates (green dots). An unbroken $E_{\nu}^{-\gamma}$ neutrino spectrum is assumed, with $\gamma = 2.0$ (left) and $\gamma = 2.5$ (right). The green line indicates the sensitivity of the combined analysis. The dashed curves indicate the sensitivities for the IceCube (blue) and ANTARES (red) individual analyses.}
\label{fig:LimitsCL}
\end{figure*}

In the candidate list search, the directions of a pre-selected list of 57 known cosmic objects, which might be promising neutrinos emitters, are investigated to look for an excess of neutrino events. The list of the analysed cosmic candidates is presented in Table~\ref{tab:LimitsFix2}, together with the results of the search. The most significant source of the list is HESSJ1023-575, a TeV $\gamma-$ray source coincident with the young stellar cluster Westerlund 2 \cite{HESSsource}. A total of 6.4 signal events and a spectral index of 3.5 are fitted at the position of the source. The trial-corrected significance of the cluster is 42\%, corresponding to 0.2$\sigma$ in the one-sided convention. Figure \ref{fig:LimitsCL} shows the upper limits at 90\% confidence level (C.L.) as a function of the source declination together with the median sensitivity, both derived with the Neyman method \cite{neyman}.

\setlength{\tabcolsep}{.3em}
\begin{table*}[p]
    \caption{\footnotesize List of cosmic objects analysed in the candidate list search. Reported are the source's name, equatorial coordinates, best-fit values of the free parameters, pre-trial p-value and 90\% C.L. upper limits on the flux normalization factor for an $E_{\nu}^{-2.0}$ spectrum, $\Phi^{90\%}_{E_{\nu}^{-2.0}}$ (in units of $10^{-9} \ \rm{GeV cm^{-2} s^{-1}}$), and for an $E_{\nu}^{-2.5}$ spectrum, $\Phi^{90\%}_{E_{\nu}^{-2.5}}$ (in units of $10^{-6} \ \rm{GeV^{1.5} cm^{-2} s^{-1}}$). Dashes (-) in the fitted number of source events, spectral index and pre-trial p-value indicate sources with null observations.
  \medskip}
    \label{tab:LimitsFix2}
    \resizebox{\linewidth}{!}{
        \begin{minipage}{1.\textwidth} 
        \centering
 {\def\arraystretch{0.85}
    \footnotesize
    \begin{tabular}{cccccccc}
   
            Name & $\delta [^\circ]$   &    $\alpha [^\circ]$    & $\hat{n}_s$  & $\hat{\gamma}$ & p-value & $\Phi^{90\%}_{E_{\nu}^{-2.0}}$ & $\Phi^{90\%}_{E_{\nu}^{-2.5}}$ \\
\hline
LHA120-N-157B & -69.16 & 84.43 & - & - & - & 3.6 & 0.9 \\ 
HESSJ1356-645 & -64.50 & 209.00 & 1.2 & 3.1 & 0.18 & 6.2 & 1.4 \\ 
PSRB1259-63 & -63.83 & 195.70 & 1.3 & 4.0 & 0.19 & 6.2 & 1.5 \\ 
HESSJ1303-631 & -63.20 & 195.74 & - & - & - & 3.7 & 0.9 \\ 
RCW86 & -62.48 & 220.68 & 1.0 & 1.6 & 0.20 & 6.3 & 1.5 \\ 
HESSJ1507-622 & -62.34 & 226.72 & - & - & - & 3.7 & 1.0 \\ 
HESSJ1458-608 & -60.88 & 224.54 & 3.7 & 3.6 & 0.036 & 9.3 & 2.0 \\ 
ESO139-G12 & -59.94 & 264.41 & - & - & - & 3.7 & 1.0 \\ 
MSH15-52 & -59.16 & 228.53 & - & - & - & 3.7 & 1.0 \\ 
HESSJ1503-582 & -58.74 & 226.46 & - & - & - & 3.7 & 1.0 \\ 
HESSJ1023-575 & -57.76 & 155.83 & 6.4 & 3.5 & 0.0079 & 11.2 & 2.5 \\ 
CirX-1 & -57.17 & 230.17 & - & - & - & 3.8 & 1.0 \\ 
SNRG327.1-01.1 & -55.08 & 238.65 & - & - & - & 3.8 & 1.0 \\ 
HESSJ1614-518 & -51.82 & 243.58 & 1.6 & 4.0 & 0.21 & 6.1 & 1.6 \\ 
HESSJ1616-508 & -50.97 & 243.97 & 2.0 & 2.0 & 0.18 & 6.5 & 1.6 \\ 
PKS2005-489 & -48.82 & 302.37 & 0.4 & 2.9 & 0.18 & 6.4 & 1.6 \\ 
GX339-4 & -48.79 & 255.70 & - & - & - & 3.7 & 1.1 \\ 
HESSJ1632-478 & -47.82 & 248.04 & - & - & - & 3.7 & 1.1 \\ 
RXJ0852.0-4622 & -46.37 & 133.00 & - & - & - & 3.7 & 1.1 \\ 
HESSJ1641-463 & -46.30 & 250.26 & - & - & - & 3.7 & 1.1 \\ 
VelaX & -45.60 & 128.75 & - & - & - & 3.6 & 1.1 \\ 
PKS0537-441 & -44.08 & 84.71 & 1.6 & 2.2 & 0.098 & 7.2 & 1.9 \\ 
CentaurusA & -43.02 & 201.36 & - & - & - & 3.6 & 1.1 \\ 
PKS1424-418 & -42.10 & 216.98 & 0.6 & 2.3 & 0.24 & 5.5 & 1.6 \\ 
RXJ1713.7-3946 & -39.75 & 258.25 & - & - & - & 3.5 & 1.2 \\ 
PKS1440-389 & -39.14 & 220.99 & 3.0 & 2.4 & 0.0085 & 10.8 & 3.0 \\ 
PKS0426-380 & -37.93 & 67.17 & - & - & - & 3.5 & 1.2 \\ 
PKS1454-354 & -35.67 & 224.36 & 3.9 & 2.1 & 0.089 & 7.3 & 2.1 \\ 
PKS0625-35 & -35.49 & 96.78 & - & - & - & 3.4 & 1.2 \\ 
TXS1714-336 & -33.70 & 259.40 & 1.2 & 2.3 & 0.17 & 5.9 & 1.9 \\ 
SwiftJ1656.3-3302 & -33.04 & 254.07 & 2.8 & 2.1 & 0.15 & 6.1 & 1.9 \\ 
PKS0548-322 & -32.27 & 87.67 & - & - & - & 3.2 & 1.2 \\ 
H2356-309 & -30.63 & 359.78 & - & - & - & 3.0 & 1.2 \\ 
PKS2155-304 & -30.22 & 329.72 & - & - & - & 3.0 & 1.2 \\ 
HESSJ1741-302 & -30.20 & 265.25 & 1.0 & 2.9 & 0.12 & 6.0 & 2.0 \\ 
PKS1622-297 & -29.90 & 246.50 & 4.4 & 1.9 & 0.048 & 7.4 & 2.4 \\ 
GalacticCentre & -29.01 & 266.42 & 2.9 & 2.1 & 0.046 & 7.2 & 2.4 \\ 
Terzan5 & -24.90 & 266.95 & - & - & - & 2.5 & 1.2 \\ 
1ES1101-232 & -23.49 & 165.91 & - & - & - & 2.4 & 1.2 \\ 
PKS0454-234 & -23.43 & 74.27 & - & - & - & 2.4 & 1.2 \\ 
W28 & -23.34 & 270.43 & 1.7 & 2.5 & 0.094 & 4.9 & 2.0 \\ 
PKS1830-211 & -21.07 & 278.42 & - & - & - & 2.2 & 1.2 \\ 
NRG015.4+00.1 & -15.47 & 274.52 & - & - & - & 1.6 & 1.0 \\ 
LS5039 & -14.83 & 276.56 & - & - & - & 1.5 & 1.0 \\ 
QSO1730-130 & -13.10 & 263.30 & - & - & - & 1.3 & 0.9 \\ 
HESSJ1826-130 & -13.01 & 276.51 & - & - & - & 1.3 & 0.8 \\ 
HESSJ1813-126 & -12.68 & 273.34 & - & - & - & 1.3 & 0.8 \\ 
1ES0347-121 & -11.99 & 57.35 & - & - & - & 1.2 & 0.8 \\ 
PKS0727-11 & -11.70 & 112.58 & 2.5 & 2.7 & 0.13 & 2.1 & 1.2 \\ 
HESSJ1828-099 & -9.99 & 277.24 & 2.4 & 2.9 & 0.079 & 2.0 & 1.2 \\ 
HESSJ1831-098 & -9.90 & 277.85 & - & - & - & 0.9 & 0.6 \\ 
HESSJ1834-087 & -8.76 & 278.69 & - & - & - & 0.8 & 0.5 \\ 
PKS1406-076 & -7.90 & 212.20 & 6.8 & 2.7 & 0.11 & 1.5 & 0.7 \\ 
QSO2022-077 & -7.60 & 306.40 & - & - & - & 0.7 & 0.4 \\ 
HESSJ1837-069 & -6.95 & 279.41 & 2.5 & 3.4 & 0.24 & 1.0 & 0.5 \\ 
2HWCJ1309-054 & -5.49 & 197.31 & 9.1 & 3.2 & 0.051 & 0.9 & 0.3 \\ 
3C279 & -5.79 & 194.05 & 2.5 & 2.2 & 0.28 & 0.6 & 0.3 \\

  \end{tabular}
      }
      \end{minipage}
      
    }
  
\end{table*}

\subsection{Galactic Centre region}\label{sec:gcs}
The third search is similar to the Southern-sky search, but restricted to a region around the Galactic Centre ($\alpha$, $\delta$) = (266.40$^\circ$,--28.94$^\circ$) and defined by an ellipse with a semi-axis of 15$^\circ$ in the direction of the galactic latitude and a semi-axis of 20$^\circ$ in galactic longitude. The results of the searches for the different investigated source extensions are presented in Table~\ref{tab:GCtable}. The most significant result is observed for an extended-source hypothesis with a width of 2.0$^\circ$ at ($\hat{\alpha}, \hat{\delta}) = (274.1^\circ, -40.1^\circ)$. The values of the best-fit $\hat{n}_s$ and $\hat{\gamma}$ are 20.3 and 3.0, respectively, and the trial-corrected significance of the hotspot is 3\% (1.9$\sigma$ in the one-sided convention).
The declination-dependent limits of this search are shown in Figure \ref{fig:GCplot}-left for different source 
extensions.

\setlength{\tabcolsep}{.3em}
\begin{table*}[h!]
    \caption{\footnotesize List of the most significant clusters found when performing the Galactic Centre region search for different source hypotheses. Reported are the extension $\sigma_s$ of the Gaussian distribution, the best-fit parameters (number of signal events, $\hat{n}_s$, spectral index, $\hat{\gamma}$,  declination, $\hat{\delta}$, right ascension, $\hat{\alpha}$), and p-values.
        \label{tab:GCtable}
    \medskip}
    \resizebox{\linewidth}{!}{
        \begin{minipage}{1.\textwidth} 
        \centering
 {\def\arraystretch{1.}
    \footnotesize
    \begin{tabular}{c|c|c|c|c|c|c}
   \hline
            $\sigma_s [^\circ]$ & $\hat{n}_s$ & $\hat{\gamma}$ & $\hat{\delta} [^\circ]$   &    $\hat{\alpha} [^\circ]$ & pre-trial p-value & post-trial p-value  \\
\hline

0.0 & 6.8 & 2.8 & -42.3 & 273.0 & $7.3 \times 10^{-4}$  & 0.40 \\ 
0.5 & 8.4 & 2.8 & -42.0 & 273.1 & $5.2 \times 10^{-4}$ & 0.19 \\ 
1.0 & 12.1 & 2.9 & -41.8 & 274.1 & $6.9 \times 10^{-4}$ & 0.15 \\ 
2.0 & 20.3 & 3.0 & -40.1 & 274.1 & $2.2 \times 10^{-4}$ & 0.03 \\ 

\end{tabular}
      }
      \end{minipage}
    }
\end{table*}

\subsection{Sagittarius A*}\label{sec:sgs}

\begin{figure*}[ht]
\centering
\begin{overpic}[width=0.45\textwidth]{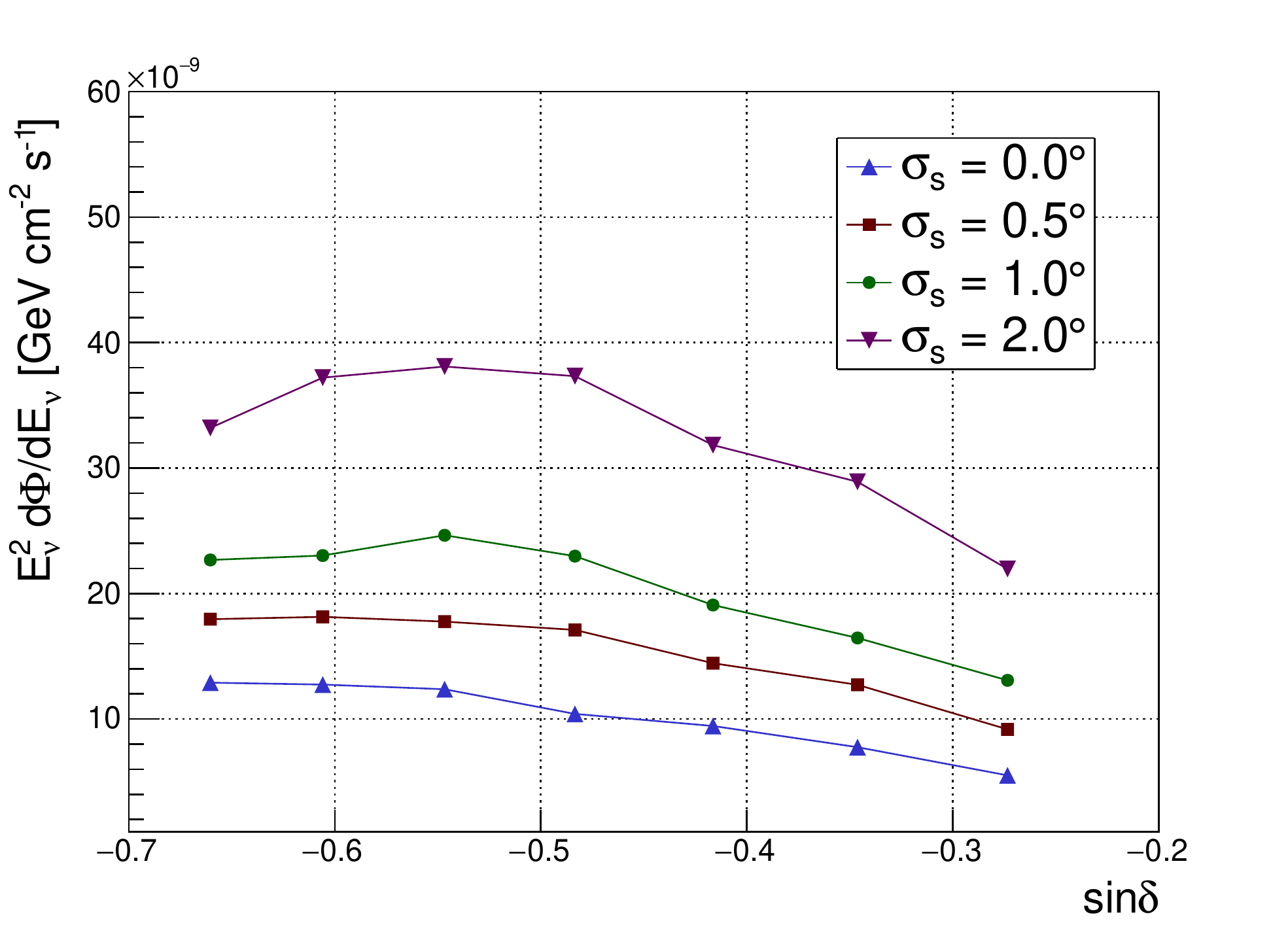}
     \put (11,10.7){\color{black} PRELIMINARY}
    \end{overpic}
\begin{overpic}[width=0.45\textwidth]{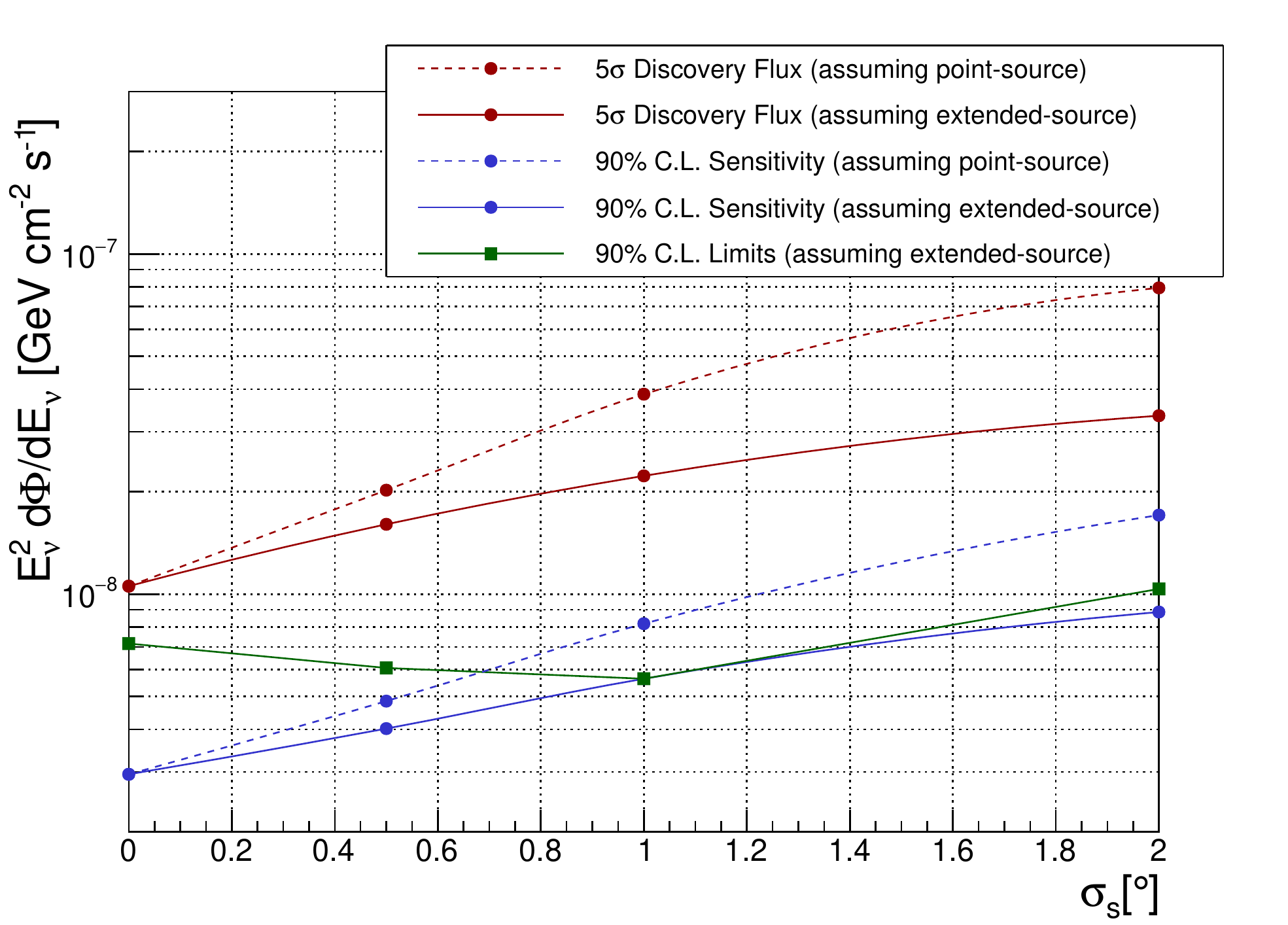}
     \put (51,10.7){\color{black} PRELIMINARY}
    \end{overpic}   
\caption{Left: 90\% C.L. limits on the neutrino flux of the Galactic Centre region search assuming an $E_{\nu}^{-2.0}$ spectrum for different source extensions $\sigma_s$. Right: discovery flux (solid red), median sensitivity (solid blue) and 90\% C.L. upper limits (solid green) for the search at the location of Sagittarius A* assuming an $E_{\nu}^{-2.0}$ spectrum as a function of the angular extension $\sigma_s$. The results for the point-like source assumption are shown as dashed lines.}
\label{fig:GCplot}
\end{figure*}

In this search, the location of Sagittarius A*, the super-massive black hole located at the centre of our Galaxy, $(\alpha, \delta) = (266.42^{\circ}, -29.01^{\circ})$, is analysed by testing point-like and extended source hypotheses. 
The values of the best-fit parameters at the investigated location for the various source widths are presented in Table~\ref{tab:SGtable} together with the p-values. The $90\%$ C.L. upper limits as a function of the source extension are shown in Figure \ref{fig:GCplot}-right together with the median sensitivity and the discovery flux.

 \setlength{\tabcolsep}{.3em}
\begin{table*}[h!]
    \caption{\footnotesize Values of the best-fit parameters (number of signal events, $\hat{n}_s$, and spectral index, $\hat{\gamma}$) and p-values for the search at the location of Sagittarius A* for different source extensions. Dashes (-) in the fitted number of source events, spectral index and pre-trial p-value indicate cases of underfluctuation.
        \label{tab:SGtable}
    \medskip}

    \resizebox{\linewidth}{!}{
        \begin{minipage}{1.\textwidth} 
        \centering
 {\def\arraystretch{0.8}
    \footnotesize
    \begin{tabular}{c|c|c|c}
   \hline
               $\sigma_s [^\circ]$ & $\hat{n}_s$ & $\hat{\gamma}$ &  p-value \\
\hline

0.0 & 2.9 & 2.1 & 0.06 \\ 
0.5 & 0.6 & 2.0 & 0.26 \\ 
1.0 & - & - & - \\ 
2.0 & 0.3 & 3.8 & 0.40 \\ 

\end{tabular}
      }
      \end{minipage}
    }
\end{table*}

\subsection{RXJ 1713.7-3946}\label{sec:rxjs}
A dedicated search for cosmic neutrinos from RXJ 1713.7-3946, the shell-type supernova remnant (SNR) at equatorial coordinates $(\alpha, \delta) = (258.25^{\circ}, -39.75^{\circ})$, is performed. Two different models are considered for the neutrino emission: the model proposed by Kappes et al. \cite{RXJ_KAPPES}, in the following indicated as RXJ 1713.7-3946 (1), and the model recently introduced for KM3NeT \cite{KM3NeTPS} and based on the methods described by Vissani et al. \cite{RXJ_VISSANI, RXJ_VISSANI2, RXJ_VISSANI3}, hereafter referred to as RXJ 1713.7-3946 (2). Both models are of the form

\small
\begin{equation} \label{eq:specRXJ}
\Phi_{\nu}(E_{\nu}) = \Phi_0 \ \left( \frac{E_\nu}{1\,{\rm TeV}} \right)^{-\Gamma} \exp (-(E_{\nu}/E_{\rm{cut}})^{\beta}),
\end{equation} 
\normalsize

\noindent where $E_{\nu}$ is the neutrino energy and the values of the neutrino spectrum parameters $\Phi_0$, $\Gamma$, $E_{\rm{cut}}$ and $\beta$ are listed in Table~\ref{tab:RXJtable}. An extension of $0.6^{\circ}$ is assumed for the source as reported by the $\gamma$-ray analysis performed by the H.E.S.S. experiment \cite{HESS}. 
No significant evidence of cosmic neutrinos from the direction of the SNR is observed for either of the considered spectra. The results of the search, in terms of number of fitted signal events and the observed p-value, are presented in Table~\ref{tab:RXJtable} for each spectrum assumption, together with the 90\% C.L. sensitivity and upper limit.

\setlength{\tabcolsep}{.3em}
\begin{table*}[h!]
    \caption{\footnotesize List of considered neutrino emission models for the search at the location of RXJ 1713.7-3946 and respective results. For each model, the values of the neutrino spectrum parameters, $\Phi_0$ (in units of $10^{-11} \rm{TeV}^{-1} \rm{cm}^{-2} \rm{s}^{-1}$), $\Gamma$, $E_{\rm{cut}}$ ([TeV]) and $\beta$, entering in equation~(\ref{eq:specRXJ}) are provided. The last four columns show the results in terms of best-fit number of signal events, $\hat{n}_s$, p-value, ratio of the sensitivity to the assumed source flux, $\Phi^{90\% \rm{C.L.}}_S/\Phi_0$, and ratio of the upper limit to the assumed source flux, $\Phi^{90\% \rm{C.L.}}_L/\Phi_0$.      
        \label{tab:RXJtable}
    \medskip}

    \resizebox{\linewidth}{!}{
        \begin{minipage}{1.\textwidth} 
        \centering
 {\def\arraystretch{1.}
    \footnotesize
    \begin{tabular}{c|c|c|c|c|c|c|c|c}
   \hline
            Spectrum & $\Phi_0$ & $\Gamma$ & $E_{\rm{cut}} $ & $\beta$ & $\hat{n}_s$ & p-value & $\Phi^{90\% \rm{C.L.}}_S/\Phi_0$ & $\Phi^{90\% \rm{C.L.}}_L/\Phi_0$ \\
\hline

RXJ 1713.7-3946 (1) & $1.55$ & 1.72 & 1.35 & 0.5 & 0.3 & 0.40 & 10.7 & 13.2\\
RXJ 1713.7-3946 (2) & $0.89$ & 2.06 & 8.04 & 1 & 0.3 & 0.41 & 9.7 & 11.7\\

\end{tabular}
      }
      \end{minipage}
    }
\end{table*}

\section{Conclusions}\label{sec:concl}
In these proceedings, the results of a combined search for neutrino sources in the Southern Sky using data from the ANTARES and IceCube detectors have been presented. No significant evidence of cosmic neutrino sources has been found in any of the different searches performed. Nevertheless, this analysis proves the high potential of the joint search for neutrino sources with ANTARES and IceCube. Thanks to the different characteristics of the experiments, the two telescopes complement each other in the search for neutrino sources in the Southern sky, allowing for a significant gain in sensitivity to possible galactic sources of neutrinos.

\vspace{10px}
\scriptsize{We gratefully acknowledge the financial support of the Ministerio de Ciencia, Innovación y Universidades: Programa Estatal de Generación de Conocmiento, ref. PGC2018-096663-B-C41 (MCIU/FEDER) and Severo Ochoa Centre of Excellence (MCIU), Spain.}

\bibliographystyle{ICRC}

\providecommand{\href}[2]{#2}\begingroup\raggedright\endgroup

\end{document}